\documentclass[12pt,preprint]{aastex}

\shorttitle{A new, kinematically anomalous H~{\small I} component in NGC~2403}
\shortauthors{Fraternali, Oosterloo, Sancisi, Van Moorsel}

\begin{document}

\title{A new, kinematically anomalous H~{\small I} component in the spiral galaxy NGC~2403}

\author{Filippo Fraternali}
\affil{Istituto di Radioastronomia (CNR), via Gobetti 101, 40129, Bologna, Italy}
\affil{Dipartimento di Astronomia, Universit\`a di Bologna, Italy}
\email{ffratern@ira.bo.cnr.it}

\author{Tom Oosterloo}
\affil{ASTRON,  P.O. Box 2, 7990 AA, Dwingeloo, The Netherlands}
\email{oosterloo@nfra.nl}

\author{Renzo Sancisi}
\affil{Osservatorio Astronomico, via Ranzani 1, 40127, Bologna, Italy} 
\affil{Kapteyn Astronomical Institute, University of Groningen, The Netherlands}
\email{sancisi@bo.astro.it}

\author{Gustaaf van Moorsel}
\affil{NRAO, P.O. Box 0, Socorro, NM 87801, USA}
\email{gvanmoor@aoc.nrao.edu}

\begin{abstract}
{
We discuss new, high sensitivity H~{\small I} observations of the spiral galaxy 
NGC~2403 which show extended emission at anomalous velocities with 
respect to the `cold' disk. This `anomalous' gas component 
($\sim$~1/10 of the total H~{\small I} mass) is probably located in the region 
of the halo and rotates more slowly ($\sim$~20$-$50 km~s$^{-1}$) than 
the gas in the disk.
Moreover, it shows a distortion in the velocity field that we interpret
as a large-scale radial motion (10$-$20 km~s$^{-1}$ inflow) towards the centre
of the galaxy.
The most likely explanation for its origin and kinematics seems to be 
that of a galactic fountain. There is, however, a significant part 
of the anomalous gas which seems to be moving contrary to rotation 
and is difficult to understand in such a picture.
These anomalous gas complexes discovered in NGC~2403 may be analogous to 
the High Velocity Clouds of our Galaxy. They may be rather common
in spiral galaxies and not have been detected yet for lack of sensitivity.
}
\end{abstract}

\keywords{galaxies: individual (NGC~2403) --- galaxies: structure ---
galaxies: kinematics and dynamics --- galaxies: halos --- galaxies: ISM}

\section{Introduction}

In recent years neutral hydrogen observations of spiral galaxies
at different inclination angles have been made with the aim of
studying the 3D distribution of the gas and the associated kinematical signature.
The study of face-on galaxies has led to the detection of vertical
motions and of `holes' in the H~{\small I} distribution that are thought to be 
produced by the expansion of large bubbles (`superbubbles') around 
stellar associations.
Detailed studies have been presented for some nearby spirals like M101 
\citep{kam93} and dwarf galaxies like Ho~II \citep{puc92}. 
The study of the edge-on galaxy NGC~891 has revealed the presence of 
neutral gas emission up to 5 kpc from the plane \citep{swa97}.
Detailed modeling of the H~{\small I} layer of this galaxy has led to the
conclusion that this emission comes from an extended H~{\small I} `halo' with a
mass of $\sim$~6$\times$10$^8$ $M_{\odot}$ (15$\%$ of the total H~{\small I} mass) 
and slower rotational velocities (by about 25 to 100 km~s$^{-1}$) 
than the gas in the disk.
Optical spectroscopy has revealed the presence of
a very thick layer (some kpc) of diffused ionized gas (DIG) in edge-on
spiral galaxies \citep{hoo99}
and also this ionized gas seems to show a vertical gradient in the 
rotation velocity. In NGC~5775 the velocity drops down to about the
systemic velocity at distances of $\sim$~5 kpc from the plane \citep{ran00}.

Thick layers of gas, large vertical motions, and rotation velocity
gradients in the vertical direction, all suggest a complex circulation of
gas, at different phases, between disk and halo of spiral galaxies.
In the classic model of galactic fountain \citep{sha76, bre80},
a continuous gas circulation is produced by supernova explosions 
and stellar winds that cause the mostly ionized gas to leave the plane 
of the disk. Once out of the plane, the ejected ionized gas is expected
to move further outwards mainly because of the pressure gradient 
and to decrease its circular velocity conserving its 
angular momentum. After cooling, clouds of neutral gas 
fall down towards the disk and acquire an inward motion.

We present here new, high sensitivity H~{\small I} observations of NGC~2403, a nearby
spiral with ongoing active star formation and the presence of bright H~{\small II} regions.
These observations bring new evidence
bearing on fountain dynamics and indeed show halo gas 
with slower rotation and give the first 
indications of a radial flow towards the centre of 
the galaxy.
However, we also detect emission from H~{\small I} which apparently 
moves contrary to rotation and is therefore difficult to place in 
the classical fountain picture.

\section{The `anomalous' gas}

\begin{figure}
\plotone{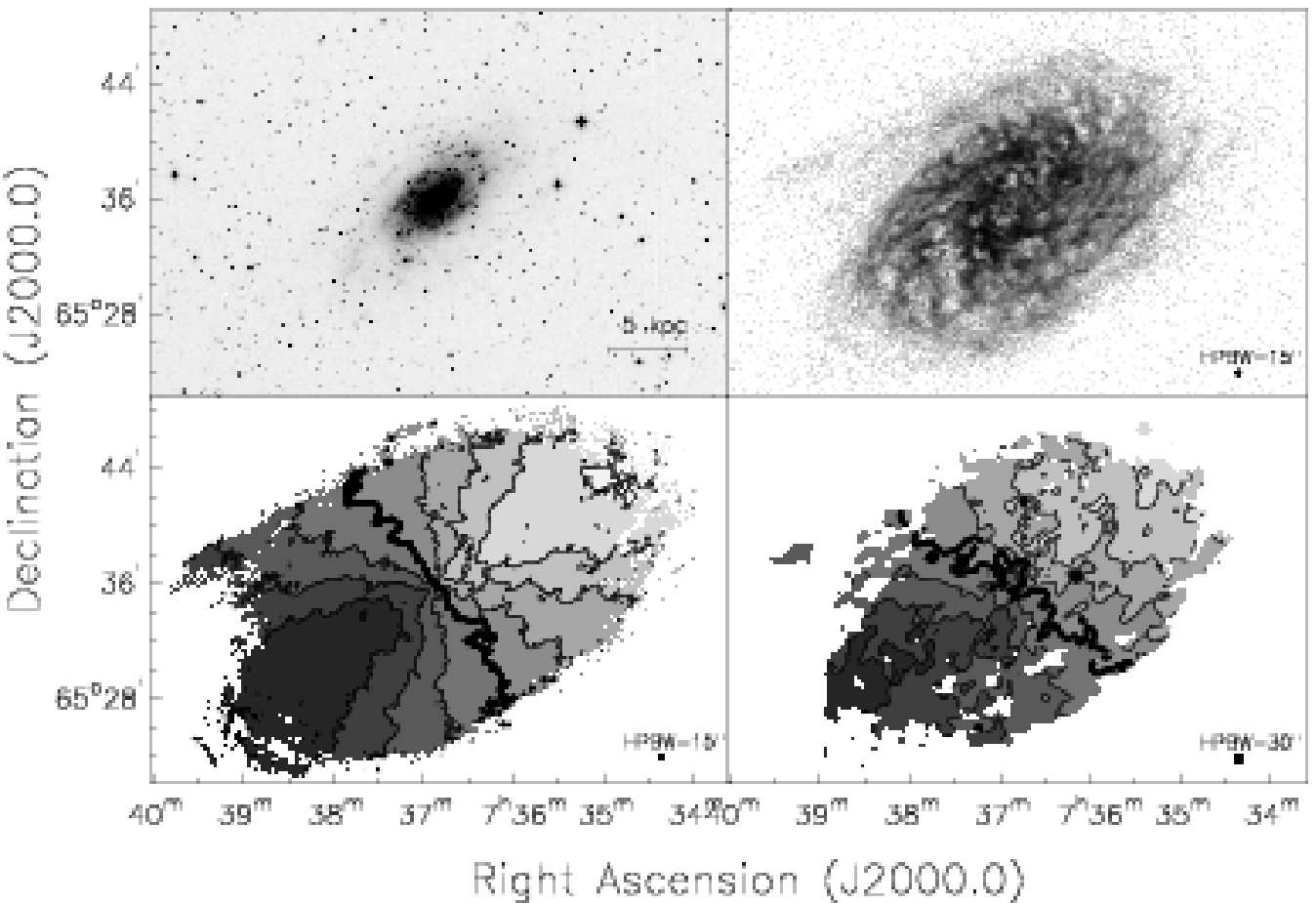}
\caption{NGC~2403. Upper panels: optical image (DSS) and total H~{\small I} density
map (VLA). The column density range in the total H~{\small I} map is
2$\times$10$^{19}-$1$\times$10$^{21}$ cm$^{-2}$. 
Bottom panels: velocity field of the cold thin disk (left) and
of the anomalous gas (right). Contour steps in both velocity fields are 30
km~s$^{-1}$. The receding side is darker, and the thick line shows the
kinematical minor axis. The systemic velocity is 133 km~s$^{-1}$.
All plots are on the same scale.
\label{fig1}}
\end{figure}

The nearby spiral galaxy NGC~2403 (Figure~\ref{fig1})
is an excellent candidate for a deep study of the density distribution and the
kinematics of H~{\small I}.
It has an extended H~{\small I} layer (the H~{\small I} radius down to a
column density of $\sim$~0.2 $M_{\odot}$pc$^{-2}$ is about 22 kpc (1$'~\simeq~$1~kpc), the
Holmberg radius is 13 kpc)
with regular kinematics and a symmetric, flat rotation curve.
Furthermore, it is viewed at an intermediate inclination angle of 60$^{\circ}$.
This offers the advantage, with respect to the edge-on and face-on views, that
information is obtained on both the motion and the density structure in the
vertical direction.

We have observed NGC~2403 with the VLA\footnote{
The National Radio Astronomy Observatory (NRAO) is a
facility of the National Science Foundation operated under cooperative agreement by
Associated Universities.}
in C configuration and a total integration time of 48 hours.
These very high sensitivity observations (the noise in the 15$''$
resolution data is 0.17 mJy/beam per channel with a velocity resolution of
10.3 km~s$^{-1}$) have revealed 
very faint H~{\small I} emission, unknown from previous observations.
Figure~\ref{fig1} shows the resulting total H~{\small I} map and velocity field;
a more detailed report of the observations will be given
elsewhere (Fraternali et al., in preparation).

The position-velocity (p-v) diagram along the major axis of NGC~2403
(Figure~2)
clearly shows systematic asymmetries in the line profiles with respect
to the rotation curve (white squares).
The profiles display faint wings of emission (we refer to them as 
the `beard') on the side of lower rotation velocities, extending 
systematically towards the systemic velocity.
Such a pattern is similar to that found in edge-on galaxies and also 
in objects observed at relatively low angular resolution.
However, NGC~2403 is sufficiently far from edge-on (i~=~60$^{\circ}$) and
our VLA observations have sufficiently high angular resolution.
It is therefore excluded that we are seeing here a component of the `cold'
disk which is projected along the line of sight because of inclination or
resolution effects.

The `beard' was already known from previous WSRT observations of NGC~2403 
obtained by \citet{sic97}, and was interpreted by \citet{sca00} as due to gas
located in the halo region of NGC~2403 and rotating 
more slowly than the gas in the plane.
They constructed 3D models of the H~{\small I} layer assuming a two-component 
structure: a thin disk and a thicker layer rotating more slowly.
Such simple models, with a mass ratio between halo and disk components
of about 15$\%$ and a difference in rotation velocity of about 25
km~s$^{-1}$, were able to reproduce the main features of the 
observed p-v diagram.

With our new data the observational picture is considerably improved
and new facts are uncovered.
The `beard' is much more extended and, completely unexpected, some emission 
also shows up in the upper right and lower left quadrants of
Figure~\ref{fig2} (forbidden for rotation).
Emission in these quadrants means non-circular motions (apparently
counter-rotating). 
We refer to the gas responsible for such emission 
as the `forbidden' gas and refer to all the gas found
at anomalous velocities (including the forbidden part) as the anomalous gas.

The total mass of the anomalous gas is $\sim$~3$\times$10$^8$ 
$M_{\odot}$ (of which just $\sim$~6$-$7$\times$10$^6$ $M_{\odot}$ forbidden), 
which corresponds to 10$\%$  of the total H~{\small I} mass of NGC~2403, and to
0.3$\%$ of its total dynamical mass.
The location on the sky of the anomalous gas can be seen (shaded areas) in
Figure~\ref{fig3} where 
we show four representative channel maps.
From them it is clear that the distribution of the 
anomalous gas is characterized by several clumpy features.
The upper left and lower right panels show the forbidden gas (arrows)
respectively in the S$-$E and the N$-$W side of the galaxy.
The other two panels, for velocities closer to systemic, show a 
remarkable coherent filament, 8 kpc long, 
with a total H~{\small I} mass of $\sim$~1$\times$10$^7$ $M_{\odot}$.

In order to study the properties of the anomalous gas we 
have separated it from the `cold' thin H~{\small I} disk by
assuming that the line profiles for the thin disk 
are represented by a Gaussian function.
We have fitted such functions at each position and subtracted them 
from the data. Similar results have been obtained by folding the 
higher velocity sides of the observed line profiles about the
profile-peak velocities (i.e. the rotation velocities) and 
subtracting them from the data cube (more details are given in
Fraternali et al., in preparation).
The mean rotation velocity derived for the anomalous H~{\small I} is about 
20$-$50 km~s$^{-1}$ lower than the disk rotation.

Figure~\ref{fig1} (bottom right) shows the velocity field for the 
anomalous H~{\small I}, to be compared with that derived for the thin
disk (bottom left).
The kinematics of the anomalous gas is clearly dominated by differential 
rotation, but its projected kinematical minor and major axes appear to be 
rotated in a counter-clockwise sense with respect to those of the 
regular disk.
The effect is more pronounced in the turn of the minor axis 
(thick line) with the result that minor and major axes are non-orthogonal.
An obvious interpretation of such a pattern is that of a non-zero radial 
component of the gas velocity.

The kinematics of the anomalous gas has been 
studied by constructing detailed models of 
the H~{\small I} layer of NGC~2403.
This has been done using the well known 
method of tilted rings and adopting a two-component structure similar 
to that used for the models of Schaap et al.\ (2000). 
The main difference is that in our 
case all the parameters 
(rotation velocity, H~{\small I} column density, position and inclination
angles)
of the  tilted ring model are derived from the data, both for the thin 
disk and for the anomalous gas.
For the velocity dispersion of the anomalous gas the data indicate 
values of 20$-$50 km~s$^{-1}$, whereas for the
cold disk the values are around 8$-$12 km~s$^{-1}$.
Figure~\ref{fig4} shows a p-v diagram parallel to the minor 
axis and centred on the major axis at 1$'$ (South-East) from the 
galaxy centre.
This position was chosen to illustrate the effects of radial motions.
The diagram shows asymmetries in the `V' shape especially visible at 
the low density levels which trace the anomalous gas.
The two models labeled with `in-flow' and `out-flow' were obtained by 
adding, for the anomalous component, a constant radial motion of $-$20 
km~s$^{-1}$ and $+$20 km~s$^{-1}$ respectively.
It is evident from Figure~\ref{fig4}  that there is a 
preference for inflow of the anomalous H~{\small I} towards the centre of the 
galaxy as opposed to an outflow or no radial motion at all.
Traces of the forbidden H~{\small I} emission, not explained by the inflow model,
are visible near the centre at velocities from 50 to 100 km~s$^{-1}$.

\section{Discussion and Conclusions}

Slowly rotating H~{\small I} halos as proposed by Swaters et al.\ (1997) for the
edge-on spiral galaxy NGC~891 and as indicated by the anomalous gas in the 
present observations of NGC~2403 may be common in spiral 
galaxies. Indeed, a similar pattern as shown here for NGC~2403 has 
also been seen in recent WSRT observations of the
spiral NGC~4559 (Oosterloo et al., in preparation).
The fact that this was unknown until recently is partly due to the poor
sensitivity of previous observations.

\begin{figure}
\plotone{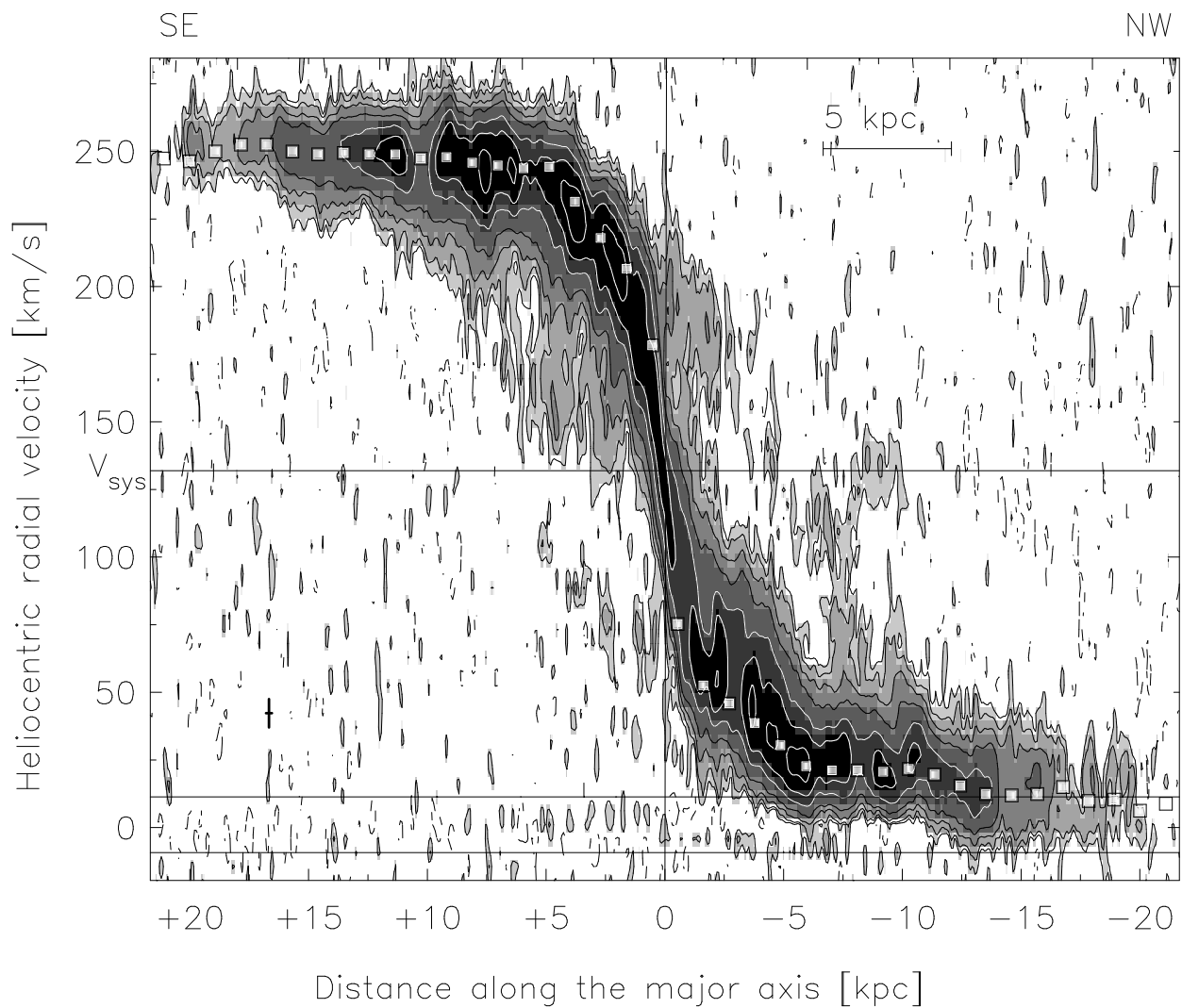}
\caption{H~{\small I} position-velocity diagram along the major axis
(p.a.~=~124$^{\circ}$, V$_{sys}$~=~133 km~s$^{-1}$) of NGC~2403 (45$''$ wide slice).
The beam size is $\sim$~15$''$.
The contours are $-$0.4, 0.4, 1, 2, 4, 10, 20, 40 mJy/beam, the r.m.s. noise is
0.17 mJy/beam.
The two horizontal lines mark the channels contaminated by emission from
our Galaxy.
The white squares show the (projected) rotation curve.
\label{fig2}}
\end{figure}

\begin{figure}
\plotone{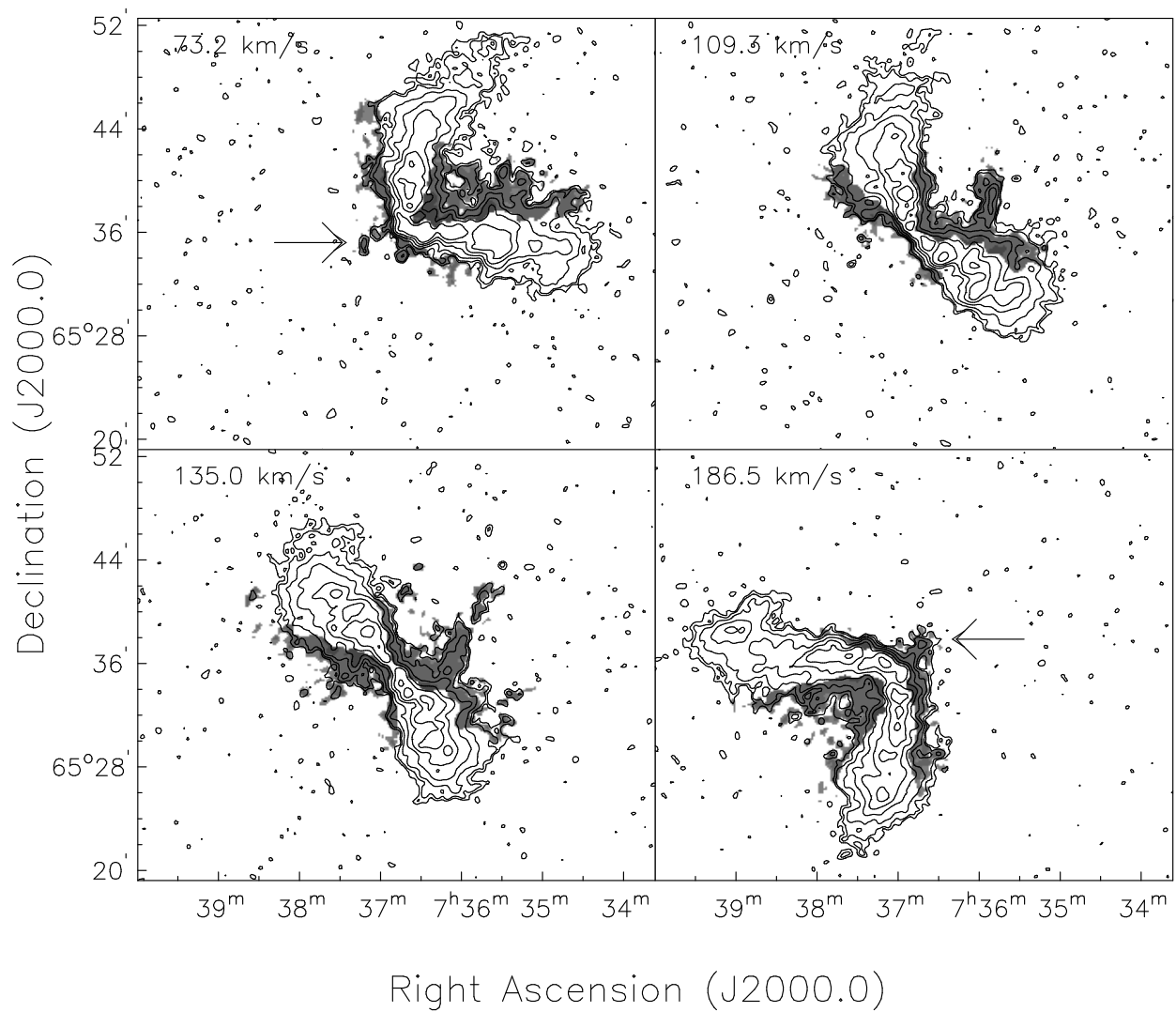}
\figcaption{Four representative H~{\small I} channel maps for NGC~2403.
The contour levels are $-$0.6, 0.6, 1, 2, 4, 10, 20, 40 mJy/beam, the
r.m.s. noise is 0.22 mJy/beam. The beam size is 30$''$.
The shading shows the anomalous gas.
The upper left and lower right panels show the forbidden gas (arrows)
while the central panels show a remarkable 8 kpc long H~{\small I} filament. 
\label{fig3}}
\end{figure}

\begin{figure}
\plotone{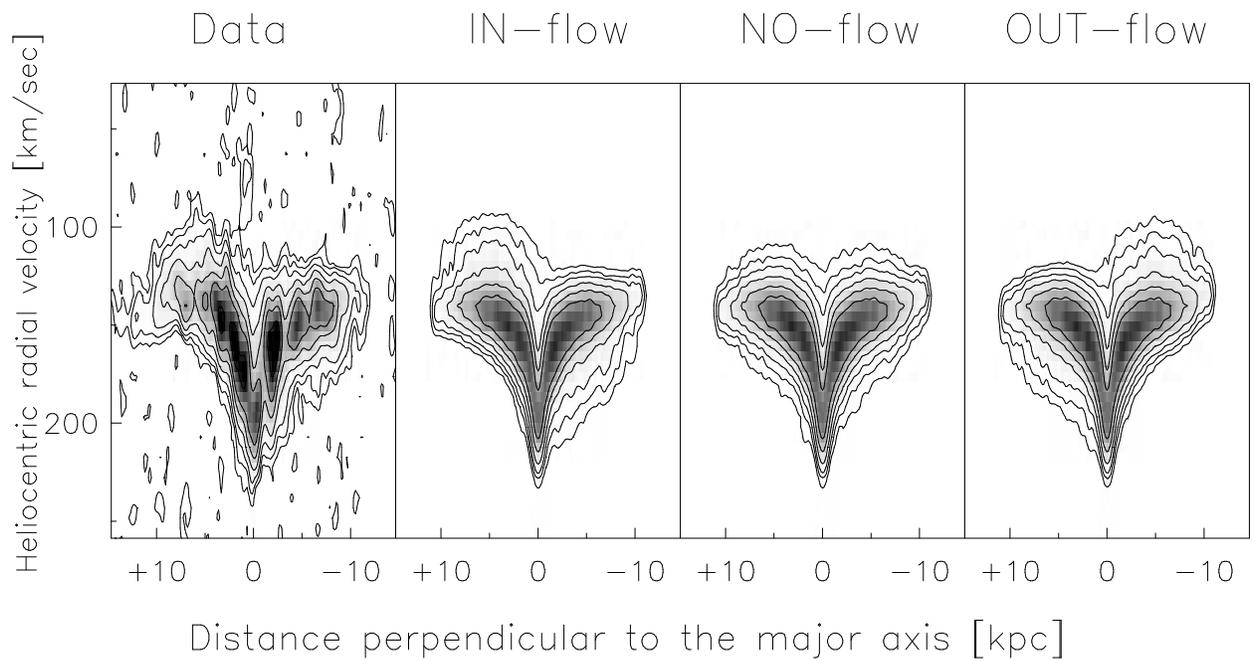}
\figcaption{Position-velocity diagram parallel to the minor
axis centred on the major axis at 1$'$ S$-$E from the centre of the galaxy.
The contour levels are $-$0.5, 0.5, 1, 2, 4, 10, 20, 40 mJy/beam, the r.m.s.
noise is 0.22 mJy/beam, and the beam size is 30$''$.
The three models are characterized by lower rotation + radial in-flow,
no-flow, and radial out-flow.
\label{fig4}}
\end{figure}

What is the origin of the anomalous gas?
Our observations of NGC~2403 have revealed two new facts. 
The first is the change in position angle of the projected
minor axis of the velocity field of the anomalous gas and 
a straightforward explanation for this is a radial flow of gas
towards the centre of the galaxy.
The second fact, probably the most surprising one, is the presence of the
forbidden gas.
This gas has projected differences from the rotation velocity of up to
150 km~s$^{-1}$ but, despite this large spread, it remains confined within 
the radial velocity range ($-$10 to $+$275 km~s$^{-1}$) 
of the galaxy rotation. Its nature is not clear but its location 
in the central bright 4 kpc of the galaxy (see Figures~\ref{fig2} and \ref{fig3})
suggests a connection with the high star formation activity.

If the picture of galactic fountain is correct, our detection of the anomalous
gas with a radial flow towards the centre of NGC~2403 may be
a direct detection of gas in the final `infalling' stage of the fountain.
There may be problems in explaining, in such a picture, long
coherent structures like the 8 kpc filament.
However, the main difficulty with a standard fountain interpretation
lies in the presence of the forbidden gas.
In order to explain that, a new approach and different
assumptions for the fountain dynamics, for instance no 
conservation of angular momentum, may be necessary.
We are now pursuing the study of this phenomenon with deep optical 
spectroscopy and X-ray {\it Chandra} observations of the central
regions of NGC~2403.

An alternative explanation for the non-orthogonality of the axes
could be that of non-circular (elliptical) orbits as expected in a triaxial
halo potential. Such a possibility is, however, not supported
by the harmonic analysis of the velocity field of NGC~2403
\citep{sch97} that suggests that the disk 
of NGC~2403 is axisymmetric to a fairly high degree.

The overall pattern and the large-scale regularity of both
the `beard' and the forbidden gas as seen in the p-v map
(Figure~\ref{fig2}) suggest that the anomalous gas forms one coherent structure.
It seems to `know' about the general pattern of disk rotation and 
to follow it closely as a broad band somewhat shifted to lower rotational
velocities. If we interpret this lower rotation
as an asymmetric drift phenomenon \citep{bin89}, 
we can estimate the velocity dispersion of the anomalous gas.
With the derived rotation curves of the disk and of the anomalous gas and
neglecting gas pressure effects we get a value for the velocity 
dispersion in the central regions of up to about 50 km~s$^{-1}$, 
not incompatible with that observed for the forbidden gas.

Finally, the anomalous gas found in NGC~2403 may be similar to
a class of the High Velocity Clouds observed in the Milky 
Way \citep{wak97}.
Although the distances and masses of HVCs are still a matter of debate
it seems fair to suggest that we have detected in NGC~2403 a population 
of H~{\small I} clouds of the same type as those intermediate and high velocity
clouds which are thought to be closely associated with the Galaxy.
The recent determination of a very low metal abundance for some of 
these clouds \citep{wak99} indicates that a part of the 
galactic high velocity gas may be {\it primordial}
\citep{oor70}. 
A similar origin
might also be advocated for the anomalous gas found in NGC~2403.
In such a picture, however, the regularity of the pattern 
(following the rotation) and the concentration of the forbidden gas 
in the optically bright (stellar) part of the system would be difficult 
to understand.

\begin{acknowledgments}
{We thank J.M. van der Hulst for helpful comments.
We acknowledge financial support (grant Cofin99-02-37) from 
the Italian Ministry for the University and Scientific 
Research (MURST).}
\end{acknowledgments}

\clearpage

\end{document}